\documentstyle[twocolumn,pra,aps,epsfig]{revtex}
\begin{document}
\draft
\title{Mutual coherence of optical and matter waves}
\author{G.A. Prataviera, E. V. Goldstein, and P. Meystre}
\address{Optical Sciences Center, University of Arizona, Tucson, AZ 85721
\\ \medskip}\author{\small\parbox{14.2cm}{\small \hspace*{3mm}
We propose a scheme to measure the cross-correlations and mutual coherence
of optical and matter fields. It relies on the combination of a matter-wave
detector operating by photoionization of the atoms and a traditional
absorption photodetector. We show that the double-detection signal is
sensitive to cross-correlation functions of light and matter waves.
\\[3pt]PACS numbers: 03.65.Bz, 03.75.-b, 42.50.-p}}
\maketitle
\narrowtext

\section{Introduction}
The manipulation and control of matter waves by optical fields forms a new
paradigm that is now receiving much attention. In a first generation of
experiments, externally applied optical fields were used to manipulate
ultracold atomic samples. Examples include optical dipole traps for
multicomponent condensates \cite{StaAndChi98} and the diffraction of 
condensates by optical gratings \cite{denHagWen99,KozDenHag99}.
More recent experiments are also concerned with the back-action
of the condensate on the light field. For example, Bragg scattering of
light off a condensate has been used to determine its form factor
\cite{SteInoChi99}, and recently, superradiance experiments
\cite{InoChiSta99} have demonstrated the joint build-up
of optical and matter-wave fields. One topic of particular interest in
this context is the possibility to generate cross-correlations and quantum
entanglement between optical and matter waves. Indeed, there are already
several theoretical predictions of such effects
\cite{MooMey98,Kua98,MooMey99,MooMey991}, e.g. in an ultracold
version of the Collective Atom Recoil Laser \cite{BonSalNar94,BonSal94}.

It is therefore timely to investigate ways to measure the joint quantum
statistical properties of optical and matter-wave fields. This is an
extension from the usual quantum optics situation, where one normally
measures
either the statistical properties of light, or more rarely, those of
the atoms, but not joined properties of the coupled
system. For this latter purpose, it is necessary to develop a generalized
detection theory applicable to the simultaneous detection of optical and
matter-wave fields.

Our approach builds on the well-known results of optical detection and
coherence theory \cite{Gla65}
and their recent extension to atomic matter waves \cite{GolMey98}.
Specifically, we consider a detection scheme consisting
of an atomic detector operating by ionization \cite{GolMey98} and a
traditional photodetector operating by absorption \cite{Gla65,CohDupGry92}.
Our model is introduced in section II. Section III presents
general expressions for the correlations between optical and matter fields.
Our general results are illustrated in section IV, where mutual coherence
functions are calculated for a simple model of linear coupling between
optical and matter waves. Finally, section V is a summary and
conclusion.

\section{The detection scheme}

We consider a situation where an optical field and an atomic matter-wave
field
are dynamically coupled, and we want to quantify the cross-correlation and
mutual coherence of these fields. The matter-wave field might, but needs
not to be, a Bose-Einstein condensate, described by a multicomponent field
operator $\bbox{\Psi}(\bf{r})$, satisfying the bosonic commutation relations
\begin{equation}
[{\Psi_i}({\bf r} ),{\Psi_j}^{\dagger}({\bf r')}]=
\delta_{ij}\delta(\bf{r}-\bf{r'}),
\label{e1}
\end{equation}
where $\bf{r}$ is the center-of-mass coordinate of the atoms and the
indices $i$ label their internal state. The optical field is described
as usual by the field operator ${\bf E}({\bf r},t)$. The dynamics of
the combined system is governed by the Hamiltonian
\begin{equation}
{\cal H}_0 ={\cal H}_s+{\cal H}_m+{\cal V}
\label{e3}
\end{equation}
where ${\cal H}_s$ and ${\cal H}_m$ describe the evolution of the
{\bf S}chr\"odinger atomic matter-wave field and of the {\bf M}axwell light
field, respectively, while  ${\cal V}$ is the interaction between these two
fields,
typically the electric dipole interaction. It is not necessary to give more
details about the specific system at hand at this point.

We now turn to the description of the detector. It consists of two parts:
The first one is a conventional optical photodetector
that emits photoelectrons when absorbing photons from the light field. In
addition, there is also a matter-wave detector. It consists of a tightly
focussed laser beam that can ionize atoms, thereby annihilating them and
producing a ion-electron pair. The detection of the mutual coherence
properties
of the coupled Maxwell-Schr\"odinger fields is obtained from the
correlations
between the electron emitted by these two detectors.

Following the original work of Glauber \cite{Gla65},
we describe the photodetector as
consisting of two-level atoms whose excited state is in the continuum,
corresponding to photoionization. We call it the {\em Maxwell detector.} Its
interaction with the system is decribed as usual by the electric dipole
interaction Hamiltonian
\begin{equation}
{\cal V}_m= -{\bf d}\cdot{\bf E}({\bf r}_m),
\label{e41}
\end{equation}
where ${\bf d}$ is the atomic electric dipole moment between the detector
ground electronic state and the continuum state, and ${{\bf r}_m}$ is the
location of the detector, assumed to be punctual for simplicity.

The interaction between the system and the {\em Schr\"odinger detector},
also assumed to be punctual at location ${\bf r}_s$, is
\begin{equation}
{\cal V}_s=\hbar \sum_i \Omega_i({\bf r}_s)
[ \Psi_i^\dagger({\bf r}_s)
 \Psi({\bf r}_s)e^{-i\omega_L t} + h.c.] .
\end{equation}
Here $ \Psi$ and $ \Psi_i$ are the field operators associated
with atoms in the sample and the continuum state
$|i\rangle $, respectively, $\Omega_i$ is a coupling constant
proportional to the electromagnetic field producing the ionization
and $\omega_L$ is the laser field frequency. The assumption of a punctual
photodetector is well justified for ultracold atomic samples, in which
case the ionizing laser can easily be focussed to a spot much smaller than
a characteristic atomic de Broglie wavelength.

The Hamiltonian describing the total interaction between the detector and
the
system is
\begin{equation}
{\cal V}_d = {\cal V}_m + {\cal V}_s ,
\end{equation}
and the mutual coherence of the optical and atomic fields is determined
to lowest order by measuring the joint counts of the Maxwell and
Schr\"odinger detectors, i.e., the joint probability to count one electron

\section{Correlations between detected signals}

To lowest order, the probability amplitude for the joint detection of
electons from the Schr\"odinger and Maxwell detectors is given by
second-order
perturbation theory in ${\cal V}_d$.
To that order, the transition probability from the system initial state
$|i\rangle$ to a final state $|f\rangle$ in a time interval $\Delta t$ is
\begin{equation}
-\frac{1}{\hbar^2}\int_t^{t+\Delta t}dt''\int_t^{t+\Delta t}dt'
\langle f|{\cal V}_d(t''){\cal V}_d(t') |i\rangle,
\label{e71}
\end{equation}
where $t'' > t'$, and the detector interaction ${\cal V}_d$ is now expressed
in the interaction picture 
\begin{equation}
{\cal V}_d(t)\rightarrow e^{i({\cal H}_0+{\cal H}_d)t}{\cal V}_d(t)
e^{-i({\cal H}_0+{\cal H}_d)t}.
\label{e8}
\end{equation}
${\cal H}_d$ being the detector Hamiltonian.

When it comes to measuring the cross-correlations between the optical and
atomic fields, the relevant terms are of course those involving
one electron each emitted by the Schr\"odinger and Maxwell detectors. In Eq.
(\ref{e71}), these are the terms involving cross-products of the 
detector-system coupling Hamiltonians ${\cal V}_s$ and ${\cal V}_m$, that
is,
\begin{eqnarray}
&-&\frac{1}{\hbar^2}\int_0^{t}dt''\int_0^{t}dt' \nonumber \\
& &\left[  \langle f|{\cal V}_m(t''){\cal V}_s(t')|i\rangle
+ \langle f|{\cal V}_s(t'') {\cal V}_m(t')|i\rangle\right ],
\label{e91}
\end{eqnarray}
with $t'' > t'$.

The probability to jointly excite the Maxwell and Schr\"odinger
detectors is therefore
\begin{eqnarray}
& &P(\Delta t)\simeq \frac{1}{\hbar ^4}\sum_{\{f\}}
{\cal R}_m(f_m) {\cal R}_s(f_s)\nonumber\\
& &\times \left |\int_t^{t+\Delta t}dt'' \int_t^{t+\Delta t}dt'
\left[\langle f|{\cal V}_m(t''){\cal V}_s(t')|i\rangle \right .\right.
 \nonumber \\
& &+ \left . \left .\langle f |{\cal V}_s(t'') {\cal V}_m(t')|i\rangle\right
]\right |^2 ,
 \;\;\;\;\; t''>t'.
\label{inter}
\end{eqnarray}
The sum is on a complete set of final states $\{f\}= \{|f_0 \rangle\otimes
| f_m\rangle \otimes |f_s\rangle \}$, where $\{|f_0\rangle \}$ is a complete
set of states of the system, and ${\cal R}_m(f_m)$ and ${\cal R}_s(f_s)$ are
the sensitivities of the Maxwell and Schr\"odinger detectors, respectively,
associated with the final states $|f_m\rangle$ and $|f_s\rangle $ of these
detectors. This sum results from the fact that we are not interested in the
final state of the system. It allows one to introduce a closure relation and
to reexpress Eq. (\ref{inter}) as
\begin{eqnarray}
& &P(\Delta t)\simeq \frac{1}{\hbar ^2} \int_t^{t+\Delta t} dt_1\int
_t^{t+\Delta t} dt_2
\int_t^{t+\Delta t} dt_3 \int_t^{t+\Delta t} dt_4 \nonumber \\
& &\times \left[\eta_s({\bf r}_s,t_4,t_3) \eta_m({\bf r}_m,t_2,t_1)\right .
\nonumber\\
& & \times
\langle E^-({\bf r_m},t_2)
\Psi^\dagger({\bf r}_s,t_4)
\Psi({\bf r}_s,t_3)E^+({\bf r}_m,t_1)\rangle 
\nonumber \\
& &+\eta_s({\bf r}_s,t_1,t_2) \eta_m({\bf r}_m,t_4,t_3)
\nonumber\\
& &\times \left .\langle
\Psi^\dagger({\bf r}_s,t_2)E^-({\bf r}_m,t_4)
E^+({\bf r_m},t_3) \Psi({\bf r}_s,t_1)\rangle \right],
\label{corr}
\end{eqnarray}
where
\begin{eqnarray}
& &\eta_s({\bf r}_s,t,t')=\frac{1}{\hbar}
e^{i\omega_{L}(t-t')}
\sum_{i}\sum_{f_s}{\cal R}_s(f_s)|\Omega_i ({\bf r}_s)|^{2} \nonumber \\
& &\times\langle i_s|
{\Psi}_{i}({\bf r}_s,t)|f_s\rangle
\langle f_s |{\Psi}_{i}^{\dagger}({\bf r}_s,t')|i_s\rangle
\label{e12}
\end{eqnarray}
and
\begin{equation}
\eta_m({\bf r}_m,t,t')=\frac{1}{\hbar}\sum_{f_m}{\cal R}_m(f_m)
\langle i_m| d (t)|f_m\rangle \langle f_m |d(t')|i_m\rangle
\label{e13}
\end{equation}
are the detector sensitivities, and $d(t)$ is the atomic dipole
operator in the interaction picture.

We have neglected contributions involving the product of two
creation and annihilation operators. This
follows from the implicit assumptions that the Maxwell detector is initially
in its ground state, and that in the Schr\"odinger field detector
any atom in the continuum will be removed from the sample instantaneously.
Note also that the statistical properties of the atoms in the detectors
do not come into play \cite{GolMey98}, a consequence of the fact that the
lowest order cross-correlations that we consider here involves only one
measurement for each of the fields involved.

Equation (\ref{corr}) can be significantly simplified in the case of
broadband detection, that is, for situations where the bandwidths of the
detectors are much broader than those of the detected fields. In this case,
and for stationary processes, the Markov approximation can be invoked in
the treatment of the detectors. Their responsivities can then be
approximated as $\delta$-functions of the time differences $t_2-t_1$ and
$t_4-t_3$.

Before proceeding with a final expression for the joint counting rates, we
observe that expression (\ref{corr}) involves the sum of two
normally-ordered correlation functions. This is a consequence of time
ordering and of the noncommutivity of the Schr\"odinger and Maxwell field
operators at different times.  One possible way to measure each of these
correlation functions separately is by using a time-gated detection scheme
\cite{CohDupGry92} with nonoverlapping in time windows for the
two detectors. Specifically, if the ionizing laser beam of the Schr\"odinger
detector is turned off after a time $t_s$  and the photodetector is
masked after $t_m$, then the coupling of the detectors to the system becomes
\begin{equation}
{\cal V} = {\cal V}_m\theta(t_m-t) + {\cal V}_s \theta(t_s-t),
\label{gate}
\end{equation}
where we can chose for example $t_s<t_m<\Delta t$. In this particular gating
scenario, only the first term in Eq.\ (\ref{corr}) remains.

Assuming that the spectrum of the Schr\"odinger and Maxwell fields to be
analyzed are centered at $\bar{\omega}_s$ and $\bar{\omega}_m$,
respectively, the probability of registering one count at each detector
during the time intervals $t_s$ and $t_m$ is then
\begin{eqnarray}
& &{\cal{P}}(t_s,t_m)=\eta_m({\bf r}_m,\bar{\omega}_m)
\eta_s ({\bf r}_s,\bar{\omega}_s)
\int_{0}^{t_m}dt_2\int_{0}^{t_s}dt_1  \nonumber \\
& &\times\langle E^{(-)}({\bf r}_m,t_2)
\Psi^{\dagger} ({\bf r}_s,t_1) \Psi({\bf r}_s,t_1)
E^{(+)}({\bf r}_m,t_2)\rangle
\label{e14}
\end{eqnarray}
where $\eta_m({\bf r}_m,\bar{\omega}_m)$ and $\eta_s ({\bf r}_s,
\bar{\omega}_s)$ are the Fourier transforms of the detector Maxwell and
Schr\"odinger responsivities. Hence the counting rate is simply propotional
to
the joint correlation function of the fields,
\begin{eqnarray}
& &w({\bf r}_s t_s,{\bf r}_m t_m)=
{\partial^{2}\over{\partial t_s\partial t_m}}{\cal{P}}(t_s,t_m)=
\nonumber \eta_m({\bf r}_m,\bar{\omega}_m)
\eta_s ({\bf r}_s,\bar{\omega}_s)\\
& &\times \langle
{E}^{(-)}({\bf r}_m,t_m)\Psi^\dagger
({\bf r}_s,t_s) \Psi({\bf r}_s,t_s)
E^{(+)}({\bf r}_m,t_m)\rangle.
\label{e16}
\end{eqnarray}

For $t_m<t_s<\Delta t$, it is only the second term in Eq.\ (\ref{corr})
that persists. It is straightforward to modify the final form (\ref{e16})
of the counting rate accordingly.

Equation (\ref{e16}) is similar to the double-photocount result in
conventional optics. This is of course not an accident, since the only
difference is that instead of having two photodetectors, we have replaced
one of them by a matter-wave detector. However, there is an essential 
difference since this new detection scheme allows one to measure joint
coherence properties of the optical and matter fields. In addition, it is
obviuous that the proposed detection scheme can readily be generalized to
measure higher-order correlatinos of these fields. In that case, though,
the fermionic quantum statistics of the ions and electrons resulting from
the
photoionization of the 
bosonic matter-wave field must be handled properly
\cite{GolMey98}. This will be the subject of a future publication.

To see how the lowest-order cross-coherence measurement of Eq. (\ref{e16})
works in practice, we illustrate it for the case of a simple example.

\section{Example}

The simplest situation one can consider is that of a linear coupling 
between the optical and matter-wave field. Of course, we should keep
in mind such a coupling is not possible in principle, due to the
conservation of the number of particles. Instead, all couplings must
be at least bilinear in the Schr\"odinger field operators, so that an
atom annihilated in one state is then created in another state. However,
the advent of Bose-Einstein condensates with a large number of
atoms in a single, macroscopically populated quantum state, allows one
to describe that state as a c-number. This is analogous to the undepleted,
classical field approximation in quantum optics. In that case, the
bilinear matter-wave part of the interaction Hamiltonian, which involves
both the condensate mode and some other side mode, approximately reduces to
a linear form only (see e.g. \cite{MooMey99,MooZobMey99}). With this kind 
of situation in mind, we then consider a matter-light interaction 
Hamiltonian of the general form
\begin{eqnarray}
{\cal H}_s&=&\sum_{\alpha=1}^N\hbar\omega_\alpha a^\dagger_\alpha a_\alpha
+\sum_{i=1}^M \hbar \omega_i c_i^\dagger c_i
\nonumber\\
&+&\hbar \sum_{i\alpha}\left[
g_i^{(1)}a_\alpha^\dagger c_i+g_i^{(2)}a_\alpha c_i+H.c.\right],
\label{lin}
\end{eqnarray}
where $N$ and $M$ are the number of modes in the optical and matter field,
respectively, with eigenfrequencies $\omega_\alpha$ and $\omega_i$.
(We use greek indices for the optical field modes and roman letters for the
matter waves in this example.) The mode annihilation operators
\begin{equation}
a_\alpha =\int d r {\bf u}_\alpha^\star({\bf r}){\bf E}^+({\bf r})
\end{equation}
and
\begin{equation}
c_i = \int d r
{\bbox\phi }_i^\star({\bf r}){\bbox \Psi}({\bf r})
\end{equation}
are bosonic optical field mode operators, ${\bf u}_\alpha({\bf r})$ and
${\bbox\phi }_i({\bf r})$ being the corresponding mode functions,
${\bf E}^+({\bf r})$ is the positive frequency part of the electric field
operator, $g_i^{(1)}, g_i^{(2)}$ are coupling constants, which allow
for a parametric amplification type of matter-light coupling as well. These
constants are proportional to the order parameter of the condensate
ground state, as discussed above.

In addition, we allow for both the optical and the matter modes to be
coupled to thermal reservoirs which result in losses at rates
$\kappa_\alpha$
and $\kappa_i$. The Langevin equations describing the evolution of
this model system are \cite{MeySar98}
\begin{equation}
\frac{d}{dt}{\bf x}(t)
 =-{\bf M}{\bf x}(t)+{\bf B}{\bbox \xi}(t),
\label{seq}
\end{equation}
where ${\bf x}(t)=\{a_\alpha(t),c_i(t)\}^T$, $\alpha=\{1,\ldots, N\};i
=\{1,\ldots, M\}$, and ${\bf M}$ is a $(N+M)\times(N+M)$ matrix whose
elements
depend on the system parameters. The diffusion matrix in Eq. (\ref{seq}) is
${\bf D}\equiv {\bf B} {\bf B}^\dagger=diag({\bar n}_\alpha\kappa_\alpha,
{\bar n}_i \kappa_i)$, where ${\bar n}_\alpha$ and ${\bar n}_i$ are the
thermal populations of the reservoir modes. We assume for simplicity that
the reservoir operators can be treated in the Markov approximation as
$\delta$-correlated white noise with zero expectation value, i.e.
$\langle \xi_i(t)\rangle=0$ and $\langle \xi_i(t) \xi_j(t')\rangle=
\delta_{ij} \delta(t-t')$.

For this example, the counting rate of Eq. (\ref{e16}) reduces to a sum
over all possible fourth-order cross-correlation functions
\begin{eqnarray}
& &w({\bf r}_s t_s,{\bf r}_m t_m)=\sum_{\alpha\beta i j}
u_\alpha^\star({\bf r}_m)u_\beta({\bf r}_m)\phi_i^\star({\bf r}_s)
\phi_j({\bf r}_s)
\nonumber\\
& &\times \left [\left( U^{-1}G(t_s,t_m)(U^{-1})^\dagger\right )_{j\alpha}
\left( U^{-1}G(t_m,t_s)(U^{-1})^\dagger\right )_{\beta i}\right .
\nonumber\\
& &+ \left .\left( U^{-1}G(t_s,t_s)(U^{-1})^\dagger\right )_{j i}
\left( U^{-1}G(t_m,t_m)(U^{-1})^\dagger\right )_{\alpha \beta }\right ],
\label{corr2}
\end{eqnarray}
where $U$ is the matrix of eigenvectors of the system matrix ${\bf M}$, so
that
\begin{equation}
U^{-1} {\bf M} U = diag(\lambda_k), k=\{1,\ldots, N+M\},
\end{equation}
where $\{\lambda_i\}$ are the eigenvalues of ${\bf M}$, and
\begin{eqnarray}
G_{ij}(t,s)&=&(U {\bf D} U^\dagger)_{ij} \frac {e^{-\lambda_i(t-s)}}
{\lambda_i+\lambda^\star_j}  \mbox{\hspace{2.cm}} t\ge s,
\nonumber\\
G_{ij}(t,s)&=&(U {\bf D} U^\dagger )_{ij} \frac {e^{-\lambda_j^\star(s-t)}}
{\lambda_i+\lambda^\star_j}  \mbox{\hspace{2.cm}} t<s .
\end{eqnarray}
The two terms in square brackets in Eq. (\ref{corr2}) correspond to
the stationary value of the counting rate and to the exponential approach
to that value. These expressions follow directly from the
 stationary solution of Eq. (\ref{seq}),
\begin{equation}
{\bf x}(t)= \int_{-\infty}^{t}
e^{-{\bf M}(t-t')}{\bf B}{\bbox \xi}(t')dt',
\label{e23}
\end{equation}
where ${\bf x}(0)=\{a_\alpha,c_i\}^T$.

The possibility to extract particular mode correlation functions from these
general expressions depends on the specifics of the problem. 
To gain some insight, we consider the two-mode problem with the drift matrix
\begin{equation}
{\bf M}=\pmatrix{\kappa_1/2& i g\cr
ig& \kappa_2/2},
\label{mat}
\end{equation}
and diffusion matrix
 \begin{equation}
{\bf D}=\pmatrix{\bar{n}_1\kappa_1&0\cr
0& \bar{n}_2\kappa_2}.
\label{dif}
\end{equation}

\begin{figure}
\epsfig{figure=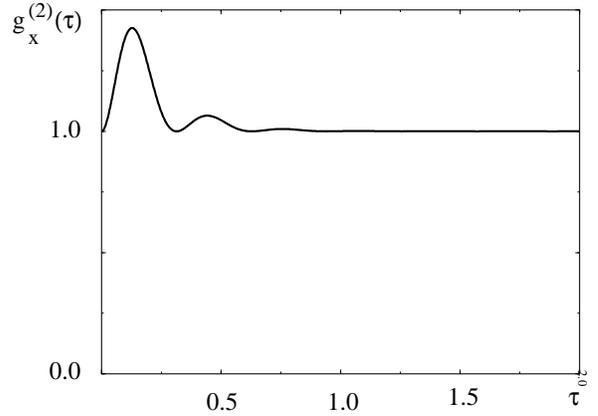,width=3.in}
\caption{
Time dependence of the normalized stationary cross-correlation
function $g_x^{(2)}(\tau)$ of the optical and matter-wave modes
for $g=10.$, $\kappa_1=6.$,
$\kappa_2=6.$, $\bar{n}_1=0.1$ and $\bar{n}_2=0.1$}
\end{figure}

The normalized stationary cross-correlation function $g_x^{(2)}(\tau)$
of the optical and matter-wave modes is shown in Fig.1 as a function of the 
time difference $\tau = t_m-t_s$. 
As expected, for this case of linear system with positive diffusion
coefficients, $g_x^{(2)}(\tau)$ exhibits bunching and 
$\lim_{\tau\rightarrow \infty} g_x^{(2)}(\tau)=1$. The oscillations correspond
to the fact that for the chosen parameters the eigenfrequencies 
of the system are complex, with a real part smaller than the imaginary 
part.

\section{Discussion and conclusions}
In this paper, we have generalized the familiar approach to optical
detection
and coherence theory to deal with situations involving coupled Maxwell and
Schr\"odinger fields. We have proposed a concrete detection scheme that
allows 
one to experimentally determine cross-correlation functions and the mutual
coherence of these fields. We have concentrated only on the lowest-order
cross-correlation, but the scheme is obviously easily generalized to
measure higher-order correlations. In that latter case, though, proper care
must
be taken of the quantum statistics of the detected signal. This topic,
where the fact that atoms are actually composite bosons (or fermions) plays
a central role, will be discussed in a future publication.

We have illustrated our detection scheme on a simple case of linear 
coupling between an atomic matter wave and an optical field. The results in 
that case are quite similar to the situation involving only optical fields. 
This is a direct consequence of the linearization procedure leading to our 
model Hamiltonian. More realistic situations will require the use of an 
interaction which is at least bilinear in the matter-wave field, in which 
case major differences can be expected, for instance in situations 
involving small condensates.

The detection scheme presented here, as well as its generalizations, are
expected to find applications in a number of situations involving e.g. the
optical control and manipulation of matter waves, nonlinear atom optics, the
generation and detection of quantum entanglement between light and matter,
and other applications of quantum atom optics. Note however that some of 
our simplifying assumptions, such as e.g. the model a punctual detector, 
only hold provided that the ionizing laser can be focused to a spot small 
compared to an atomic de Broglie wavelength. As such, it is geared toward 
ultracold samples. It would not be difficult to expand our scheme to 
detectors of finite size, but in that case, the detected photocounts will
involve a convolution with the spatial detector resolution, with a 
concomitant washing out of the finer spatial structure of the
cross-correlations.

\acknowledgments{
We are indebted to M.\ G.\ Moore for numerous discussions and valuable
suggestions. This work is supported in part by the U.S.\ Office of Naval
Research
under Contract No.\ 14-91-J1205, by the National Science Foundation under
Grant No.\ PHY98-01099, by the U.S.\ Army Research Office, and by the
Joint Services Optics Program. G. A. P. thanks Fapesp for financial
support.}


\end{document}